\documentclass[12pt,aps]{revtex4}

\usepackage{amsmath}
\usepackage{graphicx}
\usepackage{hyperref}

\begin{document}

\title{The EPR Paradox for the Uninitiated}
\author{Alexander Lesov}\affiliation{School of Physics, Georgia Institute of Technology} \email{alesov3@gatech.edu}
\date{\today}

\maketitle

Quantum mechanics (QM) is very odd. It presents both an immensely
practical and a deeply troubling conception of the physical world.
As such, its uses stretch from optimizing nanoelectronics to
examining the very nature of reality. In this article we will see
how QM forces us into a dramatic rethinking of one of the most
fundamental tenets of modern science, locality.

\section{Locality According to Special Relativity}

Locality is the idea that things can only be affected by their
immediate surroundings. By the early 20th century special
relativity had shown that nothing, including information, could
travel faster than light. This suggested a more concrete meaning
for ``immediate surroundings'': Systems could only affect one
another if the space-time gap between them was
time-like\cite{krane}. Or, in plain English, two events can only
have a cause and effect relationship if light could have travelled
from the location of the first event to that of the second in the
time between their respective occurrences.

Imagine the sun has just now blown up. The sun blowing up is an
event which happens, like every other event, at some point in
space and time. The picture below is a space-time diagram.
Obviously it's not a very accurate space-time diagram since it's
not four dimensional; never mind, we'll make do. Any event whose
location and time are known can be represented by a point on such
a diagram. As always we have freedom to choose our coordinate
system however we like. Let's make the explosion of the sun the
event which takes place at the origin, the vertex of the cones.

\begin{figure}[h]
\centering
\includegraphics[width=0.5\textwidth,height=10cm]{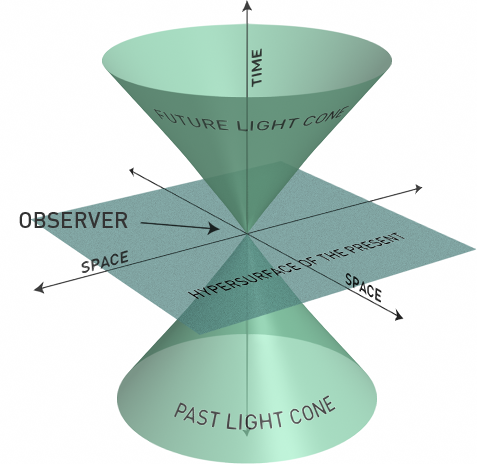}
\caption{Space-time diagram with light cones. Ignore
the``Observe'' label \cite{lightcone}.}
\end{figure}

Now let's consider another event: An observer on Earth looking up
at the sky. The observer is a certain distance and a certain
direction from the sun. In reality we would, of course, need three
dimensions to characterize his spatial location relative to the
sun but let's pretend we can do it with two. Then the observer's
space-time position at the instant of the sun's explosion lies
somewhere on the ``hyper surface of the present" (hyper because
it's three dimensional and, thus, isn't really a surface at all).
As time marches on, the space-time position of the observation
event simply travels up the time axis, with its space coordinates
remaining more or less the same (neglecting Earth's motion).
Eventually, the space-time point which represents the observation
event will fall within what is labelled the ``future light cone".
It is at this point that our unsuspecting observer's sky will
suddenly take on a particularly ominous hue. The events within the
future light cone may have been caused by the explosion of the sun
while the events within the past light cone may have been the
causes of that explosion. Anything outside of those regions can
have no causal relationship to whatever happened at the origin, in
this case, the sun blowing up. The former events, which lie within
the two cones, are called time like while the latter are called
space like. Hopefully now my original definition of locality makes
more sense. To restate it:

\begin{quote}

\textbf{Locality is the idea that two events can only be causally
related if their space-time separation is time like (i.e.they lie
within each others light cones).}

\end{quote}

By the way, the light cones can be thought of as follows: light
propagates outward from the event at the origin at a fixed speed
in every direction. Thus, at the initial moment obviously the
light hasn't had time to get anywhere. After some time the light
will have travelled all the way to the outer surface of some
sphere (represented as a circle in our two dimensional
representation of space). These circles (really spheres) are all
stacked on top of each other for each moment of time, resulting in
the cones that you see in the diagram.

Thus, the future light cone tells us how far light will have
travelled from the origin in a given time. Anything outside of
this cone couldn't possibly be caused by whatever happened at the
origin; this is because there is no way for the information
produced by the event at the origin to have reached it without
having travelled faster than light, which is impossible.
Similarly, the past light cone encloses all of the events which
could have been the causes of the event which took place at the
origin. The longer ago a causal event took place, the larger the
space it can effect at the present moment.

Now that we've introduced locality, let's figure out what quantum
mechanics has to say about it.The focus of this discussion will,
eventually, be on a 1935 thought experiment\cite{EPR} imagined by,
who else, Einstein as well as two other physicists named Podolski
and Rosen. This thought experiment lead to the deduction of what
is now called the EPR paradox.

In order to understand the EPR paradox we will consider a simple
example based on the quantum mechanical  idea of spin angular
momentum (or just spin). It would be prudent, then, to first
introduce and try to wrap our heads around some of the weird
properties of quantum spin.

\section{Quantum Spin}

Spin, in the classical sense, is nothing new: The Earth spins,
tops spin, etc. In the quantum world, however, we can't think of
spin in the same way. This is because, in QM, the particles which
we are discussing are often fundamental building blocks of matter
(e.g. electrons); fundamental in the sense that they are point
particles, indivisible and with no insides. When something spins,
in the classical sense, it rotates about some point within itself.
A point particle, though, has nothing within itself about which to
rotate!

So if the point particles don't actually spin, why give them a
property with that name? Well, although the actual motion of the
particles is certainly not one we would recognize as classical
spinning, there are subtle similarities between the classical and
quantum notions of spin. When a charged, macroscopic ball spins it
creates a magnetic field. Specifically, it becomes a magnetic
dipole with a North and a South pole. This is because of Maxwell's
Equations which state, among other things, that movement of
electrical charge (i.e. current) creates a magnetic field. Charged
quantum particles, like the electron shown below, create magnetic
fields just as if they were macroscopic balls of charge spinning
in the classical sense. Thus, we at least have some kind of
justification for calling spin, spin.

\begin{figure}[h]
\centering
\includegraphics{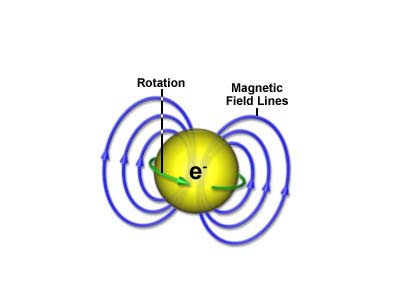}
\caption{Electron as a magnetic dipole \cite{emr}}
\end{figure}

The next weird thing about spin in QM is perhaps not very
surprising: it is quantized. We will be thinking mostly about
electrons throughout so we'll use them as our example here.
Electrons have a certain value of spin, call it a. We can set up
an experiment \footnote{A Stern-Gerlach detector which uses
inhomogeneous magnetic fields to isolate different spins can be
used for this purpose} to measure the component of the spin in a
certain direction. Our classical intuition leads us to expect that
our measurement might return values in a continuous range from -a
(spin pointing anti-parallel to axis of measurement) to +a
(parallel). In other words, we expect to get all kinds of results:
-a/4, -a/8, +7a/32, whatever. In fact, though, what we find is
that only two values are ever measured: -a and +a. Nothing in
between. That is what I meant by spin being quantized: When
measured, it can only take on one of two values instead of any
value in the continuous range from -a to a, as we would expect
classically. Bizarre? Yes. But it's what puts the quantum in
quantum mechanics.

Now that we have some idea what spin is, we can begin to develop
and understand the EPR paradox and the thought experiment that
spawned it.

\section{The EPR-Bohm Experiment}

Consider a particle sitting around doing nothing. A pion, say. Our
pion then suddenly decays to an electron and a positron (i.e.
anti-electron) in a pretty typical event. The pion has no
intrinsic spin angular momentum (this is a property of pions) and,
let's say, no other angular momentum. Armed with the knowledge
that total angular momentum must be conserved in the decay, we
deduce that the net angular momentum of the electron and the
positron must also add to zero. Now, let's set up some detectors,
one to detect the spin of the electron and one for the spin of the
positron. We'll suppose that the two detectors measure spin along
the same axis, which we call the z-axis.

Say the measurement of the electron returns the value +a. We then
immediately know, even without a direct measurement, that the spin
of the positron must be -a so that the total adds to zero. That's,
essentially, the EPR paradox. Seems a bit anticlimactic, doesn't
it? Well, it's not. In order to understand why, we need to take a
closer look at what makes QM so weird in the first place. There
are several ways to interpret the mathematical formalism of
quantum mechanics. The two that we care about here are the realist
and orthodox (AKA Copenhagen) interpretations. To introduce the
two, we'll reconsider the experiment we just thought up.

\subsection*{What Quantum Mechanics Tells Us}

Quantum mechanics is all about probabilities. It can always answer
questions like "What is the probability that the measurement of
this electron's spin will yield result x?" but cannot always
answer questions like "What is the spin of this electron?" In
other words, instead of concrete, black and white predictions, QM
often can only provide us with probabilistic answers.

So, for example, in the EPR-Bohm experiment we talked about
earlier (the one with the decaying pion) quantum mechanics cannot
tell us what the spin of the electron or the positron is until one
of them is measured by the detector. We only know that the total
spin must be zero. The two obvious possibilities are that either
the positron is spin up and the electron is spin down or the other
way around, with the electron being spin up and the positron spin
down. QM tells us that the system is actually in some combination
(i.e. superposition) of those two states. What that means
physically, don't ask me \footnote{This is the origin of the
difficulty posed by the famous problem of Schroedinger's cat
which, being dependent on a quantum mechanical event for survival,
apparently exists in a superposition of alive and dead until
observed}.

Realists have issues with this probabilistic gobbledygook
(Einstein's complaint that "God does not play dice" characterizes
their position pretty well) while followers of the Orthodox
interpretation take quantum mechanics at face value. Let's try to
understand both of these positions in more detail by attempting to
advocate each.

\subsection*{The Realist Position}

What kind of nonsense is this? If a physicist can't make 100\%
accurate, completely dependable predictions of the results of
measurements, then his physics isn't right! If quantum mechanics
can't give us concrete answers instead of some kind of
probabilistic garbage, then it must be missing something! There
must be some kind of hidden variables that haven't been taken into
account.

Here's what happens in the EPR-Bohm experiment: After the pion
decays the electron and the positron both get a spin in some well
defined direction. If we knew all the hidden variables then we
would know, without any measurement, exactly which of the two
particles has spin up and which has spin down. The fact that QM
can't tell us that kind of thing just proves that it is an
incomplete theory.

A complete theory would give us knowledge of all the variables
hidden from us by QM and, thus, allow us to predict the results of
the two measurements with total certainty. That's physics!

\subsection*{The Orthodox Position}

Quantum mechanics has never made any wrong predictions before and
has proven useful in tons of different arenas. Let's just trust it
at face value, bizarre as it might seem.

When the pion decays the electron and the positron go into a kind
of state of indecision. This state is the result of adding
(actually subtracting, but that's irrelevant) the two obvious
states we mentioned earlier:

(Superposed State) = (Electron Up, Positron Down State) +
(Electron Down, Positron Up State)

When the measurement is carried out on the system, it has to
choose between one of these two obvious states. In this case,
either has a 50-50 chance of being chosen. Once the measurement
has been made we know the state the system is in and the realists
among us can rest easy. So, for example, if we measure it to be in
the (Electron Up, Positron Down State) we can then safely say that
the electron has a spin pointing up and the positron has one
pointing down. We could not say this before the measurement,
though. In other words, we have to accept the proposition that the
measurement itself affects the system in a dramatic way.

\subsection*{The Realist's Response: The EPR Paradox}

Suppose the detectors are one light year away from one another.
Also, let's say that the electron detector is significantly closer
to the decaying pion's initial location than the positron
detector. Then it's safe to assume that the electron will reach
its detector in less time than the positron. Now imagine the
electron's detector returns the result that the electron was spin
up. If we know the electron is spin up then the positron must, at
that very same instant, be spin down in order to conserve angular
momentum.

So, according to an orthodox interpretation the fact that the
electron was measured as spin up somehow travels across one light
year of space instantaneously and causes the positron to become
spin down. This violates locality and requires faster than light
transfer of information!

\section{Summary}

That's the EPR paradox: If we believe the orthodox interpretation
of QM then locality is violated, as demonstrated in our simple
thought experiment. This implies that, in order to preserve
locality we need to put our stock in the realist interpretation,
which claims that quantum mechanics is an incomplete theory and
there are some kind of hidden variables lying around somewhere
that no one knows about.

In other words, because of the EPR paradox either locality or QM
is wrong. Both can't be right.

Experiments like the one we discussed above have convincingly
shown (see Bell's Theorem) that QM is a correct theory and that
locality is, in fact, violated. The concepts we discussed in the
previous three posts lie at the heart of tremendously exciting
recent experiments in quantum teleportation and computation which
may be the basis for all kinds of revolutionary new technologies.

\end{document}